\documentclass[twocolumn,10pt,showpacs,amsmath,amssymb]{revtex4}

\usepackage{graphicx}
\usepackage{bm}
\usepackage{epsf}
\begin{document}

\newcommand{\MNRAS}{MNRAS}
\newcommand{\ApJ}{ApJ}
\newcommand{\rhomax}{\rho^{\rm max}}
\newcommand{\gsm}{g\,cm$^{3}$}

\title{Coulomb tunneling for fusion reactions in dense matter:\\
       Path integral Monte Carlo versus mean field}

\author{A.~I.~Chugunov}
\affiliation{Ioffe Physico-Technical Institute,
Politekhnicheskaya 26, 194021 Saint-Petersburg, Russia}

\author{H.~E.~DeWitt}
\affiliation{Lawrence Livermore National Laboratory, Livermore,
CA 94550, USA}

\author{D.~G.~Yakovlev}
\affiliation{Ioffe Physico-Technical Institute,
        Politekhnicheskaya 26, 194021 Saint-Petersburg, Russia{}}

\begin{abstract}
We compare Path Integral Monte Carlo %(PIMC)
calculations by Militzer and Pollock (Phys.\ Rev.\ B {\bf 71}, 134303, 2005)
of Coulomb tunneling in nuclear reactions in dense matter
to semiclassical calculations assuming WKB Coulomb
barrier penetration through the radial mean-field potential. We find
a very good agreement of two approaches at temperatures
higher than $\sim {1 \over 5}$ of the ion plasma temperature.
We obtain a simple parameterization of the
mean field potential and of the respective reaction rates.
We analyze Gamow-peak
energies of reacting ions in
various reaction regimes and discuss
theoretical uncertainties of nuclear reaction rates taking
carbon burning in dense stellar matter as an example.
\end{abstract}

\maketitle

%%%%%%%%%%%%%%%%%%%%%%%%%%%%%%%%%%%%%%%%%%%%%%%%%%%%%%%%%%
\section{Introduction}
\label{sect-introduct}
%%%%%%%%%%%%%%%%%%%%%%%%%%%%%%%%%%%%%%%%%%%%%%%%%%%%%%%%%%

Nuclear fusion reactions in dense stellar matter affect
the evolution of ordinary stars and compact stars such as
white dwarfs and neutron stars. Hydrogen and helium burning,
and later the burning of carbon and heavier elements
\cite{clayton83} drives
an ordinary star through the main sequence and giant/red-giant
branch towards its final moments as a normal star.
%(as a presupernova or evolved giant).
Explosive burning of carbon and other elements in the cores of massive white dwarfs
triggers type Ia supernova explosions (see, e.g., \cite{hoeflich06} and references
therein). Thermonuclear burning of accreted matter in surface layers of neutron stars,
which enter compact binaries, produces type I X-ray bursts \cite{sb06}. Deeper burning
of carbon in accreting neutron stars is likely responsible for superbursts observed from
some X-ray bursters (e.g., Refs.\ \cite{cummingetal05,guptaetal06}). Even deeper burning
of accreted matter in pycnonuclear reactions in the crust of transiently accreting
neutron stars can power thermal radiation observed from neutron stars in soft X-ray
transients in quiescent states (see, e.g., Refs.\ \cite{guptaetal06,pgw06,lh07}). All in
all, nuclear fusion is important in all stars at all evolutionary stages.

It is well known that nuclear reaction rates in dense matter
are determined by astrophysical $S$-factors, which characterize
nuclear interaction of fusing atomic nuclei, and by Coulomb barrier
penetration preceding the nuclear interaction. We will mostly focus
on the Coulomb barrier penetration problem. Fusion reactions in
ordinary stars proceed in the so called classical thermonuclear
regime in which ions (atomic nuclei) constitute nearly ideal
Boltzmann gas. In this case the Coulomb barrier between reacting
nuclei is almost unaffected by plasma screening effects produced
by neighboring plasma particles. The Coulomb barrier
penetrability is then well defined.

However, in dense matter of white dwarf cores and
neutron star envelopes the ions form a strongly
non-ideal Coulomb plasma, where the plasma screening effects
are very strong. Plasma screening greatly influences the barrier penetrability
and the reaction rates. Depending on the density and temperature
of the matter, nuclear burning can proceed in four other regimes
\cite{svh69}.
They are the thermonuclear regime with strong plasma screening,
the intermediate thermo-pycnonuclear regime,
the thermally enhanced pycnonuclear regime, and
the pycnonuclear zero-temperature regime.
The reaction regimes will be briefly discussed in Sec.\ \ref{sect-regimes}.
In these four regimes the calculation of the Coulomb barrier
penetration is a complicated problem. There have been many attempts to
solve this problem using several techniques but the exact solution
is still a subject of debates. Various techniques and approaches
will also be outlined in Sec.\ \ref{sect-regimes}.

In Sec.\ \ref{sect-mp} we analyze recent Path Integral Monte Carlo (PIMC) calculations
of fusion reaction rates by Militzer and Pollock \cite{mp05}.
%AIC: removed to be same as in PhysRev
%%%DG included reference
%%(also see their
%%preliminary results \cite{pm04}).
%% using the Path Integral Monte Carlo
%%(PIMC) technique.
We compare these results
with those obtained within
a much simpler formalism of semi-classical Coulomb tunneling in a
mean-field potential. To this aim,
in Sec.\ \ref{sect-mfwkb} we analyze and parameterize
the mean field potential
in a strongly non-ideal classical ion plasma and also
calculate and parameterize mean-field
reaction rates. Section \ref{sect-analysis}
is devoted to comparison of calculations by
different authors. We conclude in Sec.\ \ref{sect-conclusions}.

%%%%%%%%%%%%%%%%%%%%%%%%%%%%%%%%%%%%%%%%%%%%%%%%
\section{NUCLEAR REACTION REGIMES}
\label{sect-regimes}
%%%%%%%%%%%%%%%%%%%%%%%%%%%%%%%%%%%%%%%%%%%%%%%%

%%%%%%%%%%%%%%%%%%%%%%%%%%%%%%%%%%%%%%%%%%%%%%%%%%%%%%%%%%%%%%%%
\subsection{Physical parameters}
\label{sect-physics}
%%%%%%%%%%%%%%%%%%%%%%%%%%%%%%%%%%%%%%%%%%%%%%%%%%%%%%%%%%%%%%%%

\begin{figure}
    \begin{center}
        \leavevmode
        \epsfysize=3.2in \epsfbox{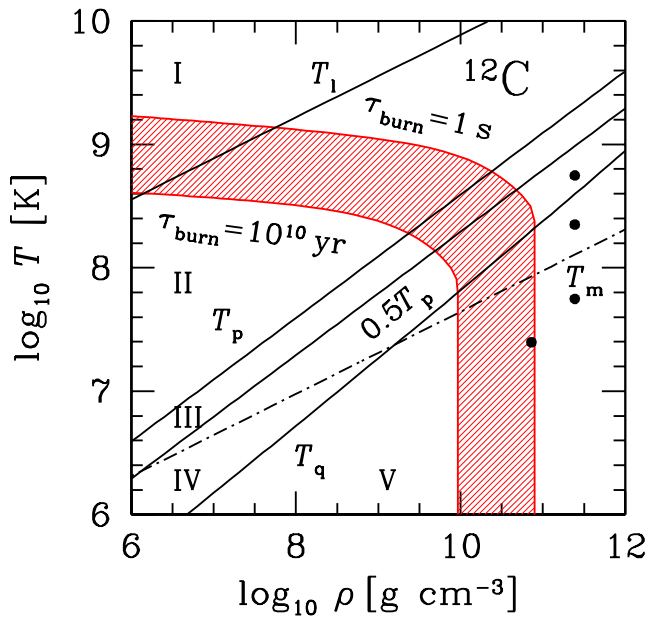}%{diagram.ps}
    \end{center}
    \caption{(Color online)
    Temperature-density diagram for carbon matter;
    $T_l$ ($\Gamma=1$) is the temperature below which
    ions form strongly coupled liquid; $T_p$ ($\zeta\approx 0.513$)
    is the ion plasma temperature; $T_m$ is the solidification
    temperature in a classical ion liquid; $T_q$ is the temperature
    below which the carbon burning rate is temperature independent;
    I--V label domains of different nuclear burning regimes
    (Table \ref{tab:regimes}).
    The shaded region is most important
    for applications of
    carbon burning; it is restricted by the lines of constant
    burning times (1~s and $10^{10}$ yr, for the upper and lower
    lines, respectively, from Ref.\ \cite{leandro05}).
    Filled dots show some $T-\rho$
    points for which PIMC calculations \cite{mp05}
    of reaction rates have been
    performed if applied to carbon burning.
    }
    \label{fig:diag}
\end{figure}

We consider a fusion reaction between identical nuclei
$(A,Z) + (A,Z)$ in a one-component plasma (OCP) of atomic nuclei
(ions) in dense matter. Here, $A$ and $Z$ are
the mass and charge numbers of the nuclei, respectively.
The results will be general
but they will be illustrated taking the $^{12}$C+$^{12}$C
reaction as an example that is most important for
astrophysical implications (Sec.\ \ref{sect-introduct}).
The temperature-density ($T - \rho$) diagram for carbon matter is shown in
Fig.\ \ref{fig:diag}. The filled dots show some
$T-\rho$ points for which PIMC calculations
\cite{mp05} have been performed if applied to carbon burning.
Notice that the majority of the data points \cite{mp05}
correspond to much higher $T$ and $\rho$, where
carbon would be immediately transformed into other
elements either through beta captures or through intense
nuclear reactions. All the PIMC data \cite{mp05}
are analyzed in Secs.\ \ref{sect-mp}--\ref{sect-analysis}.
The shaded region in Fig.\ \ref{fig:diag}
is briefly described in Sec.\ \ref{sect-all}.

Under the conditions
displayed in Fig.\ \ref{fig:diag},
carbon
is fully ionized
%(either by high temperature and/or by electron pressure)
and immersed in a nearly uniform electron background;
the electrons are mostly strongly degenerate.

Coulomb coupling of ions is determined by the familiar
parameter
\begin{equation}
  \Gamma  = Z^2e^2/(k_BaT ),
\label{Gamma}
\end{equation}
where
%
%\begin{equation}
     $a = [3/(4\pi n_i)]^{1/3}$
%\label{a}
%\end{equation}
%
is the ion-sphere radius, $n_i$ is the ion number density, and $k_B$ the Boltzmann
constant. At temperatures $T \gg T_l = Z^2e^2/(k_B a)$ (i.e., at $\Gamma  \ll 1$) the
ions constitute an almost ideal Boltzmann gas. At $T \lesssim T_l$ ($\Gamma  \gtrsim 1$)
they smoothly (without any phase transition) transform into a strongly coupled liquid.
The ions solidify at $T = T_m$ which is much lower than $T_l$. The freezing into the
classical body-centered cubic (bcc) crystal occurs at $\Gamma_m \approx 175$, that is at
$T_m = T_l/\Gamma_m$. The difference of free energies of various Coulomb structures at
low temperatures is very small, and the actual microstructure at these temperatures is
rather uncertain (as discussed, e.g., in \cite{leandro05,mcp06}).

The importance of quantum effects in motion of plasma ions can be
characterized by the ion plasma temperature $T_p$ determined by the
ion plasma frequency $\omega_p$,
\begin{equation}
   T_p = {\hbar \omega_p / k_B}, \quad
   \omega_p=\sqrt{4 \pi Z^2 e^2 n_i / m_i},
\label{Tp}
\end{equation}
where $m_i$ is the ion mass. At $T \gtrsim T_p$ quantum effects
are relatively weak and quantization of ion motion is mainly
unimportant. At lower $T$ quantum effects become most essential.
At $\rho \gtrsim 4 \times 10^{11}$ g~cm$^{-3}$
dense matter contains free neutrons dripped off atomic nuclei.

OCP of ions can also be characterized by the parameters
\begin{equation}
    r_s = {a \over a_B} ,
    \quad \eta = {\Gamma \over r_s},
    \quad
    \zeta=\left(\frac{4\Gamma^2}{\pi^2 \,r_s}\right)^{1/3}=
    \left( 4 T_p^2 \over 3 \pi^2 T^2 \right)^{1/3},
\label{rs}
\end{equation}
where $a_B = \hbar^2/(Z^2e^2m_i)$ is the ion Bohr radius.
A state of ions is determined by two parameters, for instance, by
$\Gamma$  and $T /T_p$ (or $\zeta$), while other parameters can be expressed
through these ones.

We will also need a temperature (Fig.\ \ref{fig:diag})
\begin{equation}
    T_q =0.5\, T_p/\ln(T_l/T_p),
\label{Tq}
\end{equation}
which is the upper temperature
of temperature-independent pycnonuclear
burning (Sec.\ \ref{sect-zeropyc}).

\begin{table}[t]
\caption[]{Nuclear reaction regimes in dense matter \cite{svh69}.}
\label{tab:regimes}
\begin{center}
\begin{tabular}{clc}
\hline
%\hline
Line& $\qquad\qquad$ Regime & Domain \\
\hline
%\hline
I & Thermonuclear with weak screening & $T \gg T_l$ \\
II& Thermonuclear with strong screening &
                              $T_p \lesssim T \lesssim T_l$ \\
III & Thermo-pycnonuclear & $0.5 T_p \lesssim T \lesssim T_p $ \\
IV  & Thermally enhanced pycnonuclear & $T_q \lesssim T \lesssim 0.5 T_p$\\
V   & $T=0$ pycnonuclear  & $ T \lesssim T_q $ \\
\hline
\end{tabular}
\end{center}
\end{table}

%As already mentioned in Sec.\ \ref{sect-introduct},
The five nuclear reaction regimes
%(Salpeter and Van Horn
\cite{svh69}
%)
%in dense matter. They
are summarized in Table \ref{tab:regimes}.
Temperature-density domains for carbon burning in these regimes
are seen from Fig.\ \ref{fig:diag}.
We outline these regimes for fusion reactions
of identical nuclei in an OCP.

%%%%%%%%%%%%%%%%%%%%%%%%%%%%%%%%%%%%%%%%%%%%%%%%%%%%%%%%%%%%
\subsection{Thermonuclear burning with weak plasma screening}
\label{sect-TW}
%%%%%%%%%%%%%%%%%%%%%%%%%%%%%%%%%%%%%%%%%%%%%%%%%%%%%%%%%%%%

This regime (regime I in Table \ref{tab:regimes})
is realized at $T \gg T_l$.
In this case Coulomb coupling of ions is weak
and Coulomb tunneling is only slightly
affected by plasma screening effects. The main
contribution to reaction rate comes from a small
amount of ions with energies $E$ near the Gamow-peak
energy $E_\mathrm{pk}$ (much higher than $k_B T$).

Neglecting the plasma screening effects,
one obtains the familiar
%classical expression for the
thermonuclear reaction rate
\begin{equation}
    R_\mathrm{th}= 4\,{n_i^2 \over 2} \,
    \sqrt{ 2 E_\mathrm{pk} \over 3 \mu} \,
    { S(E_\mathrm{pk}) \over k_B T } \, \exp(-\tau),
\label{therm}
\end{equation}
where $E_\mathrm{pk}=k_B T \tau /3$, $\mu=m_i/2$
is the reduced mass, $S(E)$ is the astrophysical factor
(assumed to be a slowly varying function of $E$), and
\begin{equation}
  \tau= \left( 27 \pi^2 \mu Z^4 e^4 \over
        2 k_B T \hbar^2 \right)^{1/3}.
\label{tau}
\end{equation}

The plasma screening enhances the reaction rates. In the
thermonuclear regimes with weak and strong screening
(at $T \gtrsim T_p$, and actually even at
somewhat lower $T$) the ``screened'' rate can be conveniently written
as
\begin{equation}
    R^\mathrm{scr}_\mathrm{th}=R_\mathrm{th}\,F_\mathrm{scr},
    \quad F_\mathrm{scr}=\exp(h),
\label{Fscr}
\end{equation}
where $R_\mathrm{th}$ is given by Eq.\ (\ref{therm}) and
$F_\mathrm{scr}$ is the enhancement factor expressed
through a function $h$.
In the weak screening Debye-H\"uckel limit ($\Gamma \ll 1$, $T \gg T_l$),
one obtains \cite{salpeter54}
$h = \sqrt{3}\,\Gamma^{3/2} \ll 1$, so that
 $F_\mathrm{scr}\approx 1+h$
is only slightly higher than~1.

%%%%%%%%%%%%%%%%%%%%%%%%%%%%%%%%%%%%%%%%%%%%%%%%%%%%%%%%%%%%%%%
\subsection{Thermonuclear burning with strong plasma screening}
\label{sect-TS}
%%%%%%%%%%%%%%%%%%%%%%%%%%%%%%%%%%%%%%%%%%%%%%%%%%%%%%%%%%%%%%%

This regime (regime II in Table \ref{tab:regimes}) occurs at
$T_p \lesssim T \lesssim T_l$.
The majority of ions are strongly coupled by Coulomb
forces in their potential wells; quantum effects in
their motion are weak. The main contribution
to the reaction rate comes from a small amount of
highly energetic ions which
are nearly free and have Gamow-peak energies (modified by
the screening effects).
The plasma screening strongly
enhances the reaction rate.

The screening is often modeled assuming that
the reacting nuclei move in a potential
\begin{equation}
  U(r)=Z^2e^2/r-H(r),
\label{W(r)}
\end{equation}
where $H(r)$ is a static and spherically
symmetric
mean-field plasma potential.
This approach neglects fluctuations of plasma
screening microfields during
an individual tunneling event.

Generally, the function $h$ can be split into two parts,
\begin{equation}
   h=h_0+h_1, \quad F_\mathrm{scr}=\exp(h_0) \exp(h_1).
\label{h0h1}
\end{equation}
The leading term $h_0=H(0)/k_BT$ is calculated assuming a constant mean-field plasma
potential $H(r)=H(0)$ during the quantum tunneling, while $h_1$ is a correction owing to
a weak variation of $H(r)$ along the tunneling path and owing to possible deviations
from the mean-field approximation (concerned with fluctuations). Estimates show (e.g.,
Ref.\ \cite{ys89}) that
%Yakovlev and Shalybkov 1989),
typical tunneling
lengths of the reacting ions in the thermonuclear regime
($T  \gtrsim T_p$) are smaller than the ion sphere radius
$a$, and typical tunneling times are shorter than the plasma
oscillation time scales $\sim \omega_p^{-1}$.
This justifies the approach
of almost constant and static plasma potential during a tunneling
event as a first approximation.

The main screening quantity $h_0$ is determined by $H(0)$. For a classical ion system,
$H(0)$ can be calculated as $H(0)=\Delta {\cal F}$, where $\Delta {\cal F}$ is a
difference of the Coulomb free energy for a given system of nuclei and for a system with
two nuclei merged into one compound nucleus (e.g., DeWitt \textit{et al.}\
\cite{dgc73}). In this case the leading enhancement factor $\exp (h_0)= \exp( \Delta
{\cal F}/k_BT)$ depends on the  one argument $\Gamma$.

The mean-field potential $H(r)$ for a classical strongly coupled OCP of ions (liquid or
solid) can be determined from classical Monte Carlo (MC) sampling (e.g., DeWitt {\it et
al.}\ \cite{dgc73}). We will analyze the latest results in Sec.\ \ref{sect-mfpot}. MC
sampling gives the static classical radial pair distribution function of ions $g(r)$
which equals $g(r)=\exp[-U(r)/k_BT]$. In this way one obtains accurate values of $g(r)$
and $H(r)=Z^2e^2/r+k_BT \ln g(r)$ at not too small $r$ (typically, at $r \gtrsim a$),
because MC statistics of close ion separations $r \lesssim a$ is poor due to strong
Coulomb repulsion of ions at small distances. The potential $H(r)$ at small $r$,
required for calculating the tunneling probability, can be obtained by extrapolating MC
values of $H(r)$ to $r \to 0$. The extrapolation is a delicate procedure (as shown,
e.g., by Rosenfeld \cite{rosenfeld96}).

Assuming a linear mixing rule in a
multi-component strongly coupled ion plasma, Jancovici
\cite{jancovici77} obtained
\begin{equation}
    h_0(\Gamma)=2f_0(\Gamma)-f_0(2^{5/3}\Gamma),
\label{h0f0}
\end{equation}
where $f_0(\Gamma)$ is a Coulomb free energy per one ion in an
OCP (in units of $k_B T$). Using MC data available by
that time (1977)
he got $h_0(\Gamma)$ given in line (a) of Table \ref{tab:h0}.
The same expression was used by Itoh {\it et al}.\ \cite{ikm90}.

\begin{table*}[t]
\caption[]{Function $h_0(\Gamma)$ as calculated by different authors.}
\label{tab:h0}
\begin{center}
\begin{tabular}{llll}
\hline
%\hline
Line& ~~~Ref.~~~  & ~~~~~~~~~~~~~~~~~~~$h_0(\Gamma)$~~  &~~ $\Gamma$~~~ \\
\hline
%\hline
(a)&
Eq.\ (17) in \cite{jancovici77} &
$1.0531\,\Gamma+2.2931\,\Gamma^{1/4}-0.5551\,\ln\Gamma-2.35$ &
$1 \leq \Gamma \leq 155$ \\
(b)~~~ & %[best]&
Eq.\ (20) in \cite{leandro05}~~~~ &
$1.0563 \, \Gamma + 1.0208\, \Gamma^{0.3231}-0.2748 \,
\ln \Gamma -1.0843 $~~~~ &
$1 \leq \Gamma \leq 170$ \\
(c)&
Eq.\ (6) in \cite{oii91} &
$1.148 \,\Gamma-0.00944\,\Gamma \ln \Gamma-0.000168\,\Gamma (\ln \Gamma)^2 $ &
$5 \leq \Gamma \lesssim 180$ \\
(d) &
Eq.\ (19) in \cite{ogata97} &
$1.132\,\Gamma-0.0094\,\Gamma \ln \Gamma $ &
$1 \lesssim \Gamma \lesssim 170$ \\
\hline
%\hline
%\scriptsize{$^{*)}$Best}  & & & \\
\end{tabular}
\end{center}
%&\scriptsize{ $^{*)}$ Most accurate up to now.}~~~~~~~~~~~~~~~~~~~~~~~~~~~~~~~~~~~~~~~~~~~~
\end{table*}

Recent MC
% AIC: removed, to be the same as in PhysRev
%%DGY inserted hypernet chain
%%and hypernet chain
calculations for a classical Coulomb liquid at $\Gamma
\gtrsim 1$
% AIC: to be the same as in paper
give highly accurate values of $f_0(\Gamma)$ (accurately approximated by analytical
functions, e.g., Ref.\ \cite{pc00}) and confirm the validity of the linear mixing rule
(e.g., Ref.\ \cite{ds03}).
%give highly accurate values of $f_0(\Gamma)$ and confirm the validity of the linear
%%DGY inserted dsc96
%mixing rule (e.g., Refs.\ \cite{dsc96,ds03}).
%%mixing rule (e.g., Refs.\ \cite{pc00,ds03}).
The function $h_0(\Gamma)$ has been calculated from Eq.\ (\ref{h0f0}) in many papers
(e.g., \cite{jancovici77,ys89,rosenfeld96,ds99}), and the results are in good agreement.
In line (b) of Table \ref{tab:h0} we present an analytical approximation of
$h_0(\Gamma)$, which follows from the recent MC results of DeWitt and Slattery
\cite{ds99} for a Coulomb liquid. It seems to be the best available evaluation of
$h_0(\Gamma)$, which is in very good agreement with the expression in line (a). In the
indicated $\Gamma$-range, it is accurately approximated by a linear function
$h_0(\Gamma)\approx 0.9\,(2^{5/3}-2)\,\Gamma = 1.0573\,\Gamma$ suggested by Salpeter
\cite{salpeter54} using a simple ion-sphere model. At $\Gamma \gg 1$ the plasma
screening enhancement is huge. For instance, $\exp(h_0) \sim 10^{74}$ for $\Gamma \sim
170$.

Some authors
calculated $H(0)$ and the related enhancement factor
$\exp(h_0)$ by
extrapolating MC $H(r)$ to $r \to 0$ (as mentioned above).
In particular, Ogata {\it et al}.\ \cite{oii91,oiv93} used that formalism
to analyze the enhancement of nuclear reactions
in one-component and two-component
strongly coupled ion liquids.
Their $h_0(\Gamma)$ is given in line (c) of Table \ref{tab:h0}.
These calculations
are less accurate than those based on Eq.\ (\ref{h0f0}) because
of the problems of extrapolation of $H(r)$ to $r \to 0$
in Refs.\ \cite{oii91,oiv93} (see
Refs.\ \cite{rosenfeld96,leandro05,mcp06}
for details).

The function $h_0(\Gamma)$ was also calculated by
Ogata \cite{ogata97} using
the PIMC method. His result [line (d) in Table \ref{tab:h0}]
is in better
agreement with the most accurate result [line (b)] %Eq.\ (\ref{accfit})
as discussed in Ref.\ \cite{leandro05}.

In addition to $\exp(h_0)$, the enhancement factor $F_\mathrm{scr}$ in Eq.\ (\ref{h0h1})
contains a factor $\exp(h_1)$ which depends on two arguments (e.g., $\Gamma$ and
$\zeta$). The basic term in $h_1$ in the thermonuclear regime with strong screening was
obtained by Jancovici \cite{jancovici77}; it is presented in line (A) of Table
\ref{tab:h1}, where $\zeta$ is given by Eq.\ (\ref{rs}). Introducing $\tau$ from Eq.\
(\ref{tau}) we have $\zeta=3 \Gamma /\tau \approx r_t/a \sim (T_p/T)^{2/3}$, $r_t$ being
the Coulomb tunneling length in the thermonuclear regime ($T \gtrsim T_p$). Clearly,
$\zeta$ can be regarded as a small parameter in that regime. This basic term can be
easily obtained in the mean-field approximation (Sec.\ \ref{sect-enhancement}) with the
lowest-order expansion terms of $H(r)$ over $r$, $H(r)=H(0)-{1\over 4}\,(r/a)^2\,
Z^2e^2/a$. Treating the $r^2$ correction as small and using the semi-classical
approximation for the tunneling probability, one immediately comes to Eq.\ (A).

\begin{table*}[t]
\caption[]{Function $h_1(\Gamma,\zeta)$ as calculated by different authors.}
\label{tab:h1}
\begin{center}
\begin{tabular}{llllll}
\hline
%\hline
Line& ~~~Ref.~~~  & ~~~~~~~~~~~~~~~~~~~$h_1(\Gamma,\zeta)$~~
&~~ $\Gamma$~~~&~~$\zeta$~~&~~$T/T_p$ \\
\hline
%\hline
(A)&
Eq.\ (35) in \cite{jancovici77} &
$-(5/32)\,\Gamma\,\zeta^2    $ &
$1 \leq \Gamma \leq 155$ & $\zeta \lesssim 1$ & $T/T_p \gtrsim 0.37$ \\
(B)~~& %[best]&
Eq.\ (28) in \cite{aj78} &
$-(5/32)\,\Gamma\,\zeta^2+ 0.014 \Gamma \zeta^3+0.0128\, \Gamma \zeta^4 $ &
$1 \leq \Gamma \leq 155$ & $\zeta \leq 1.6$ & $T/T_p>0.18$ \\
(C)& Eq.\ (36) in \cite{oii91}~~~& $-(5/32)\,\Gamma\,\zeta^2
\left[1+\left(1.1858-0.2472\,\log\Gamma\right)\,\zeta-0.07009\zeta^2\right]$~~~~
&
$5 \leq \Gamma \lesssim 180$~~~ & $\zeta \lesssim 2$~~~ & $T/T_p\gtrsim 0.13$\\
(D) &
Eq.\ (19) in \cite{ogata97} &
$-(5/32)\,\Gamma\,\zeta^2 (1-0.0348\zeta-0.1388\zeta^2+0.0222\zeta^3) $ &
$1 \lesssim \Gamma \lesssim 170$ & $\zeta \lesssim 2$ & $T/T_p\gtrsim 0.13$ \\
\hline
%\hline
  &
\end{tabular}
\end{center}
%$^{*)}$ Most accurate up to now.\hwidth
\end{table*}

Equation (A) in Table \ref{tab:h1} can be treated as the
well defined lowest-order term in the expansion
of $h_1(\Gamma,\zeta)$ in powers of $\zeta$.
There were several attempts to improve
Eq.\ (A) by adding new terms
obtained either on theoretical grounds or by fitting
numerical results. These new terms are model dependent
and debatable. It is thought that adding
these terms allows one to extend the results to lower temperatures,
somewhat beyond the lowest boundary $T \sim T_p$
for the thermonuclear regime (let us remark that $T=T_p$
corresponds to $\zeta=0.513$, and
$T=T_p/2$ corresponds to $\zeta=0.815$). We will discuss the validity
of such extensions in Sec.\ \ref{sect-analysis}.

Alastuey and Jancovici \cite{aj78} proposed semi-analytic
corrections to (A). %beyond the
%mean-field approximation.
Their result is given in line
(B) of Table \ref{tab:h1}.
Ogata {\it et al}.\ \cite{oii91,oiv93}
calculated the Coulomb tunneling probability
and $h_1(\Gamma,\zeta)$ using the mean-field
potential and solving numerically an effective
radial Schr\"odinger equation.
A fit to their calculations is given in line (C).
These results were used by Kitamura \cite{kitamura00}
for constructing an analytic expression for
nuclear reaction rates in all burning regimes.
Ogata \cite{ogata97} calculated $h_1(\Gamma,\zeta)$ using
PIMC. His fit is presented in line (D).
Itoh {\it et al}.\ \cite{ikm90} determined the enhancement factors
$F_\mathrm{scr}$
calculating the WKB Coulomb barrier penetrability in a mean-field
potential. Their results are
equivalent to
$h_1(\Gamma,\zeta)=1.25\,\Gamma -\tau \,f(\zeta)-h_0(\Gamma)$,
where $h_0(\Gamma)$ is given by Eq.\ (a) of Table \ref{tab:h0}
and $f(\zeta)$ is given by their lengthy fit expression (4.4) [with
our $\zeta$ denoted by $\beta$ and $\zeta\leq 5.4$].
Let us remark that their mean-field potential $H(r)$ is simplified.
Its $H(0)$ value is correct but small-$r$ behavior
is approximate (a continuous function with a break).

Recent PIMC calculations by
Militzer and Pollock \cite{mp05}
will be analyzed in Sec.~\ref{sect-analysis}.

%%%%%%%%%%%%%%%%%%%%%%%%%%%%%%%%%%%%%%%%%%%%%%%%%%%%%%
\subsection{Zero-temperature pycnonuclear burning}
\label{sect-zeropyc}
%%%%%%%%%%%%%%%%%%%%%%%%%%%%%%%%%%%%%%%%%%%%%%%%%%%%%%

Zero-temperature pycnonuclear regime (regime V in Table
\ref{tab:regimes}) takes place
at low temperatures, $T \lesssim T_q$,
particularly, at $T=0$; $T_q$ is defined by
Eq.\ (\ref{Tq}) and plotted in Fig.\ \ref{fig:diag}. In this case, one can
safely assume that all plasma ions occupy ground
states in their potential wells. Quantum tunneling and
nuclear fusion occur mainly between
close neighbors owing to zero-point ion vibrations;
thermal effects in ion motion are unimportant
and the reaction rate is temperature-independent.
The reaction rate increases with growing $\rho$
because zero-point vibrations become more efficient.

Coulomb tunneling probability in this regime has
been calculated in various approximations
(see Refs.\ \cite{leandro05,mcp06} for a recent analysis
of the results).
The predicted reaction rates have similar density dependence
but differ within several orders
of magnitude.
%New accurate calculations are required
%(Sec.\ \ref{sect-pycno-mp}).

%%%%%%%%%%%%%%%%%%%%%%%%%%%%%%%%%%%%%%%%%%%%%%%%%%%%%%%%%%%%
\subsection{Thermally enhanced pycnonuclear burning}
\label{sect-thermoexpyc}
%%%%%%%%%%%%%%%%%%%%%%%%%%%%%%%%%%%%%%%%%%%%%%%%%%%%%%%%%%%%

Thermally enhanced pycnonuclear regime
(regime IV from Table \ref{tab:regimes}) occurs at higher temperatures,
$T_q \lesssim T \lesssim 0.5\,T_p$.
These temperatures are still so low
that the majority of ions occupy their
ground states. However, some of them populate higher
bound energy levels in the potential wells. These ions
give the major contribution to the reaction rate
because it is much easier for them to penetrate through
the Coulomb barrier.

This burning regime was studied
%by Salpeter and Van Horn
in Refs.\ \cite{svh69,ki95}.
As discussed in Ref.\ \cite{leandro05}, the results
differ within several orders of magnitude and have to be
improved.

When the density grows up, the upper and lower
boundary temperatures
for this regime, $0.5T_p$ and $T_q$,
become closer (Fig.\ \ref{fig:diag}) and
finally merge. Therefore, this reaction regime
disappears at sufficiently high densities (when
spacings $\sim \hbar \omega_p$
between quantum energy levels in
potential wells become too large). As a rule,
such a density is too high to be of practical importance
(it would be $\sim 10^{15}$ g~cm$^{-3}$
for carbon burning, in which case no carbon can survive
in dense matter).

%%%%%%%%%%%%%%%%%%%%%%%%%%%%%%%%%%%%%%%%%%%%%%%%%%%%%
\subsection{Thermo-pycnonuclear burning}
\label{sect-thermopyc}
%%%%%%%%%%%%%%%%%%%%%%%%%%%%%%%%%%%%%%%%%%%%%%%%%%%%%

This
regime (regime III from Table \ref{tab:regimes}) takes place
at
$T_p/2 \lesssim T \lesssim T_p$.
It is intermediate between thermonuclear regimes and
pycnonuclear ones. When the temperature increases from
$\sim T_p/2$ to $\sim T_p$, those ions, which give the most important
contribution into the reaction rate, become unbound and
move to continuum states (from closest-neighbor
collisions in the pycnonuclear case to collisions of
freely moving particles in the thermonuclear case).
%This regime is most complicated for analytical studies.

%%%%%%%%%%%%%%%%%%%%%%%%%%%%%%%%%%%%%%%%%%%%%%%%%%%%%%%
\subsection{All regimes and problems}
\label{sect-all}
%%%%%%%%%%%%%%%%%%%%%%%%%%%%%%%%%%%%%%%%%%%%%%%%%%%%%%%

Several works \cite{kitamura00,leandro05,mcp06} have suggested
analytic fits for the nuclear reaction rates
valid in all five burning regimes. In particular,
the fits constructed in Refs.\ \cite{leandro05,mcp06}
take into account current theoretical uncertainties
of the reaction rates and give the optimal, maximum and
minimum theoretical rates.

In Fig.\ \ref{fig:diag} we present the lines which divide the $T-\rho$ diagram into five
regions appropriate to five regimes. In addition, we plot two lines \cite{leandro05}
along which the characteristic carbon burning times $\tau_\mathrm{burn}=n_i/R$ are equal
to 1~s and 10$^{10}$ years (the upper and lower lines, respectively). The astrophysical
factor $S(E)$ for the carbon burning is taken from Ref.\ \cite{leandro05}. (We neglect a
possible hindrance of the carbon reaction \cite{jiangetal07} at energies much lower than
the Coulomb barrier energy.) The lines are almost horizontal at lower $\rho$ and higher
$T$, where carbon burns in thermonuclear regimes, and they are almost vertical at higher
$\rho$ and lower $T$, where carbon burns in pycnonuclear regimes. Above and to the right
of the $\tau_\mathrm{burn}=1$~s line the burning is very fast, and there is no carbon in
dense matter. Below and to the left of the $\tau_\mathrm{burn}=10^{10}$~yr line the
burning is extremely slow (practically absent). Therefore, the density-temperature
domain of practical interest for carbon burning is located between these two lines.

Although the main features of the Coulomb tunneling
in dense matter seem clear, some important
tunneling problems in OCP are still unsolved:
\begin{description}
\item{(i)} In the thermonuclear regime, where the plasma screening
is conveniently described by the enhancement factor
$F_\mathrm{scr}$ [Eqs.\ (\ref{Fscr}) and (\ref{h0h1})],
the function $h_0(\Gamma)$ is well defined [line (b) of Table \ref{tab:h0}].
But what is an exact form of the function $h_1(\Gamma,\zeta)$
(Table \ref{tab:h1})?
\item{(ii)} Down to which temperatures this description can be extended?
\item{(iii)} What is an exact expression for the reaction rate at lower
temperatures when the burning becomes pycnonuclear?
\end{description}
Our aim will be to analyse problems (i) and (ii) and discuss (iii).
The problems in multicomponent
ion mixtures are more complicated \cite{mcp06}.

%%%%%%%%%%%%%%%%%%%%%%%%%%%%%%%%%%%%%%%%%%%%%%%%%%
\section{Militzer-Pollock PIMC calculations}
\label{sect-mp}
%%%%%%%%%%%%%%%%%%%%%%%%%%%%%%%%%%%%%%%%%%%%%%%%%%

New PIMC calculations of Militzer and Pollock \cite{mp05}
were performed for a wide range of plasma parameters
which cover all regimes of nuclear burning.
In Fig.\ \ref{fig:diag}
filled dots show some densities and temperatures
of stellar matter which correspond to four Militzer-Pollock
points (the values of $\Gamma$ and $r_s$ in their
Table 1) if applied for carbon burning.
The majority of their points are not shown; they would refer
to much higher $T$ and $\rho$ than those displayed
in Fig.\ \ref{fig:diag}
(far from the $T-\rho$ region where carbon can exist in
dense stellar matter).

In principle, the PIMC is the best method to calculate Coulomb
tunneling in nuclear reactions. It can take into account all
the effects of dense environment on Coulomb tunneling,
including fluctuative nature of plasma potential;
dynamical response of plasma ions to the motion of reacting
nuclei in the course of quantum tunneling;
finite width of trajectories of the tunneling nuclei
(that is beyond
the WKB approach).
Unfortunately, highly accurate PIMC calculations
require huge computer resources (long PIMC runs with
many plasma ions involved) while all PIMC calculations
performed so far are naturally limited, at least by not
too many plasma ions.

Notice that the PIMC simulations \cite{mp05} neglected the effects of quantum statistics
of ions. As shown by several authors (e.g., Ref.\  \cite{ogata97}), this is a good
approximation for the conditions of practical interest.
%AIC: removed, to be the same as in PhysRev
%%DG
%%The same conclusions can be made from test PIMC calculations
%%of Pollock and Militzer \cite{pm04} with spin polarized particles.
Usually, nuclear burning occurs
when $\rho$ is still insufficiently high
for the quantum statistics effects to be pronounced.

PIMC simulations of Militzer and Pollock \cite{mp05}
included 54 plasma ions.
The authors
calculated the contact probabilities
$g(0)$, which are the values of the
quantum-mechanical radial pair distribution function $g(r)$
at $r \to 0$. For OCP of ions, $g(0)$ is
related to the reaction rate through \cite{ichimaru93}
\begin{equation}
\label{klevo}
  R=\frac{n_i^2}{\pi}\, \frac{a_B}{\hbar}\, S(E_{\rm pk})\, g(0),
\end{equation}
where $a_B=\hbar^2/(m_i Z^2 e^2)$.

\begin{figure}[t!]
    \begin{center}
        \leavevmode
        \epsfxsize=3.3in
    \epsfbox{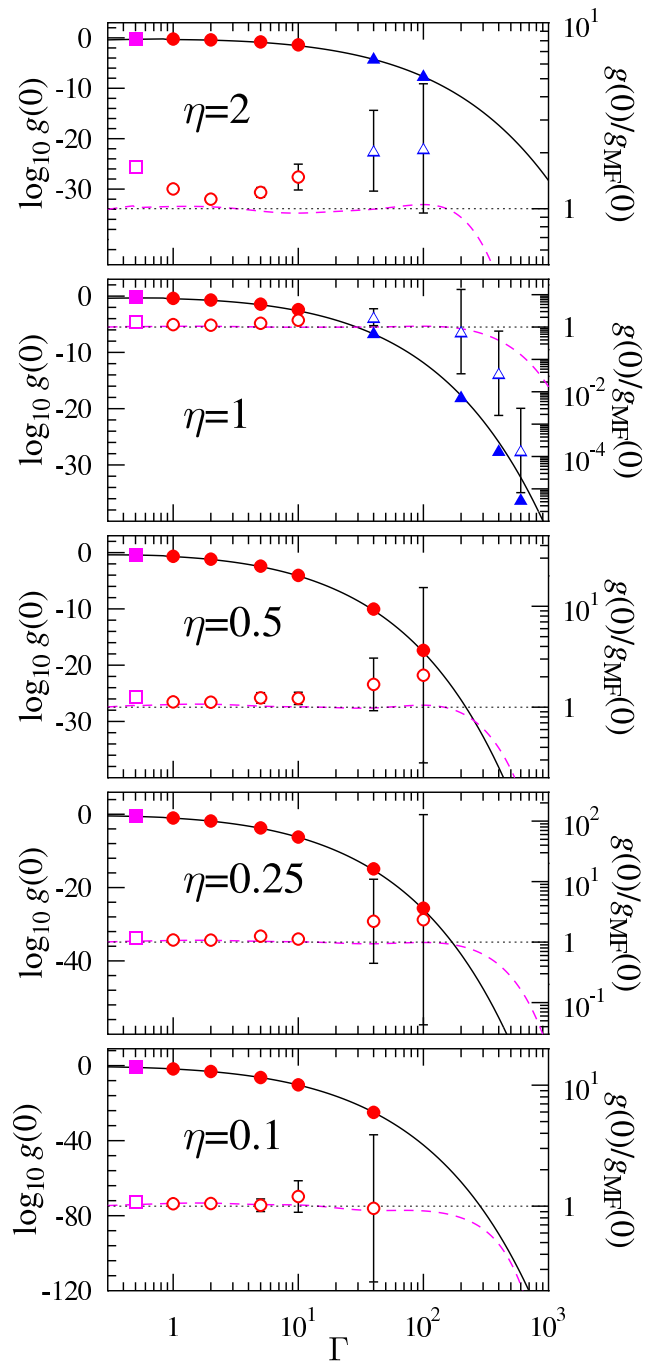}
    %{EtaGamma5_corrected_logErr.eps}
    \end{center}
    \caption{Contact probabilities
    versus $\Gamma$ for five
    values of $\eta$=0.1, 0.25, 0.5, 1, and 2.
    {\em Left vertical axis}:
    Filled symbols show all Militzer-Pollock $g_{MP}(0)$ data;
    squares, dots and triangles
    mark the data of the three groups (i)--(iii)
    (Table \ref{tab:mp}).
    Solid lines are the contact probabilities $g_\mathrm{MF}(0)$
    calculated in the mean-field WKB approximation.
   % (Sec.\ \ref{sect-mfwkb}).
    {\em Right vertical axis}: Open symbols with errorbars
    display
    the ratio
    $g_{MP}(0)/g_\mathrm{MF}(0)$
    of the PIMC to mean-field WKB results [dotted lines
    refer to $g_{MP}(0)=g_\mathrm{MF}(0)$ to guide the eye].
    Dashed lines show the ratio
    $g_\mathrm{MF}^\mathrm{fit}(0)/g_\mathrm{MF}(0)$
    of our mean-field WKB fitted (Sec.\ \ref{sect-fit})
    and calculated values.
    See text for details.
    }
    \label{fig:etagamma}
\end{figure}

Militzer and Pollock present their $g(0)\equiv g_\mathit{MP}(0)$
for 36 values of plasma parameters $\Gamma$ and $\eta$
(equivalently, for 36 values of $\rho$ and $T$, four of
which are shown in Fig.\ \ref{fig:diag}).
Specifically, they considered 10 values of $\Gamma$:
$\Gamma=0.5$
(five $\eta$-points), $\Gamma=$1 (five points), $\Gamma=$2 (five points),
$\Gamma=$5 (five points),
$\Gamma=$10 (five points), $\Gamma=$40 (five points),
$\Gamma=$100 (three points), $\Gamma=$200 (one point),
$\Gamma=$400 (one point), and $\Gamma=$600 (one point).
All their results are shown in Fig.\ \ref{fig:etagamma},
where we compare them with our calculations
in the mean-field WKB approximation
[$g(0)=g_\mathrm{MF}(0)$; see Secs.\ \ref{sect-mfwkb}
and \ref{sect-analysis} below].
Note five typos in Table I of Ref.\ \cite{mp05};
in the values of $-\ln[g(0)]$ in column 5 for
$\eta=0.25$, $\Gamma=100$; $\eta=0.5$, $\Gamma=100$;
and $\eta=1$, $\Gamma$=200, 400 and 600, one should
remove zero after dot; this removal restores correct errobars
of $-\ln[g(0)]$ (otherwise the errorbars are ten times
smaller than their actual values).

We have divided the Militzer-Pollock
data (somewhat arbitrarily) into three groups (i), (ii) and (iii)
(listed in Table \ref{tab:mp}).

\begin{table}[t]
\caption[]{Three groups of Militzer-Pollock data points \cite{mp05}.} \label{tab:mp}
\begin{center}
\begin{tabular}{clc}
\hline
%\hline
Line& $\qquad\qquad$ Group & Domain \\
\hline
%\hline
(i) & Moderate plasma screening & $\Gamma =0.5$ \\
(ii) & $T$ dependent rate, strong screening &
                              $\Gamma\ge 1,~\zeta<3$ \\
(iii) & $T$ independent pycnonuclear data & (see text) \\
\hline
\end{tabular}
\end{center}
\end{table}

The first group (i) consists of five points
with $\Gamma=0.5$ (marked by squares
in Fig.\ \ref{fig:etagamma}).
These points correspond to a moderately strong Coulomb
coupling ($T=2T_l$, thermonuclear burning intermediate
between weak and strong plasma screening regimes). We do not analyze
them in detail because in this case the effects of plasma screening
on Coulomb tunneling are weak and exact screening enhancement
factors have not been calculated so far by other theoretical
methods.

The second group (ii)
includes 26 data points (dots
in Fig.\ \ref{fig:etagamma})
with $1 \leq \Gamma \leq 100$ and
$\zeta \leq 3$ ($T \ge 0.071 T_p$).
Three such points belong
to the thermonuclear regime with strong screening ($T\geq T_p$),
seven points correspond to the intermediate thermo-pycnonuclear
regime ($0.5 T_p < T < T_p$), while other 16 points
refer to pycnonuclear burning at not too low $T$
($0.071 T_p \leq T \leq 0.5 T_p$).

The third group (iii) includes six points
(triangles in Fig.\ \ref{fig:etagamma}). Three of them
have very large $\Gamma=200$, 400 and 600 (with $\zeta=4.33$,
5.45, and 6.24, respectively). The other three have
lower $\Gamma=40$, 40, and 100 but large $\zeta$
(2.53, 3.18, and 4.32), i.e., small $T \leq 0.0912 T_p$
at which the contact probability is expected to become
temperature independent. The point with $\Gamma=40$ and
$\zeta=2.53$ belongs also to the second group.

%%%%%%%%%%%%%%%%%%%%%%%%%%%%%%%%%%%%%%%%%%%%%%%%%%%%%%%%%%%%%%%%%%%%%%%%%
\section{Mean Field WKB Approximation}
\label{sect-mfwkb}
%%%%%%%%%%%%%%%%%%%%%%%%%%%%%%%%%%%%%%%%%%%%%%%%%%%%%%%%%%%%%%%%%%%%%%%%%

In this section we calculate the nuclear reaction rate
in a mean-field WKB approximation, which is much simpler
than the PIMC.

%%%%%%%%%%%%%%%%%%%%%%%%%%%%%%%%%%%%%%%%%%%%%%%%%%%%%%%%%%%%%%%%%%%%%%%%%
\subsection{Mean-field potential}
\label{sect-mfpot}
%%%%%%%%%%%%%%%%%%%%%%%%%%%%%%%%%%%%%%%%%%%%%%%%%%%%%%%%%%%%%%%%%%%%%%%%%

%
\begin{figure}[t]
    \begin{center}
        \leavevmode
        \epsfxsize=3.2in
    \epsfbox[20 25 275 335]{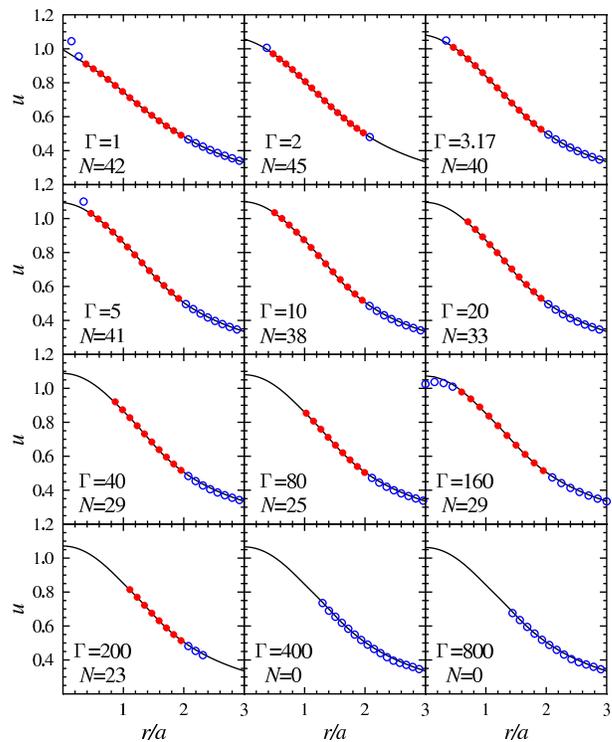}
    \end{center}
    \caption{
    (Color online) Normalized mean field MC potential
    $u(x)$ in an OCP of ions
    for 12 values of $\Gamma$ from 1 to 800.
    The data for $\Gamma \leq 200$ are obtained
    for ion liquid, while the data for $\Gamma=400$ and
    800 correspond to bcc ion crystal.
    Filled and open dots
    show data points included into fitting and excluded from
    it, respectively. Lines show the fit; $N$ is the number of data
    points included into the fitting. See text for details.
    }
    \label{fig:potential}
\end{figure}

We start with the discussion of the mean field potential
$H(r)$ in a strongly coupled OCP. We have taken the results
of extensive Monte Carlo (MC) calculations of $H(r)$
for a classical OCP of ions in a liquid phase for ten values
of $\Gamma$ from 1 to 200 within some intervals of $x=r/a$
(Fig.\ \ref{fig:potential}).
Although a thermodynamically stable phase of OCP at $\Gamma>175$
is a crystalline bcc lattice, the difference of free energies
of the lattice and liquid is small. As a result, the OCP can
stay liquid at temperatures much below the melting point,
and one can simulate supercooled liquid in MC runs.
In order to check our results we have also taken
MC calculations of $H(r)$ for a classical bcc Coulomb
crystal at $\Gamma=400$ and 800 (two last panels in
Fig.\  \ref{fig:potential}).

It is useful to introduce a dimensionless mean-field
potential $u(r)$,
\begin{equation}
     H(r)/k_B T = \Gamma u(x), %\quad x=r/a,
\label{dimensionless}
\end{equation}
which is plotted in Fig.\ \ref{fig:potential}.

The dimensionless potential $u(x)$ depends on two variables, $x$ and $\Gamma$. For
further use, we have fitted the calculated values of $u(x)$ by an analytical expression.
The fitting included selected MC points (denoted by filled dots) for an OCP liquid at $1
\leq \Gamma \leq 200$. The number $N$ of these data points at any $\Gamma$ is presented
in Fig.\ \ref{fig:potential} (for a better visualization we have not plotted some
points). Other MC points, not included into the fitting, are denoted by open dots. In
particular, we did not fit the data for the ion crystal ($\Gamma=400$ and 800). We have
also excluded from the fitting several data points at low $x$ for $\Gamma \leq 5$ and
$\Gamma=160$. These points were calculated with large errors because of poor MC
statistics at low separations $x$. We do not show their errorbars to simplify the
figure. The errorbars for all other points are small and would be invisible. The data
points at sufficiently large $x$ for $\Gamma \leq 200$, excluded from the fitting, have
been used for checking the quality of our fit. Similar potentials for OCP liquid have
been computed, e.g., in Ref.\ \cite{cg03}. The potentials  suggested in Ref.\
\cite{itw03} for $4 \leq \Gamma \leq 90$ and $x \lesssim 1.6$ are also very accurate.

To ensure a high fit accuracy at small $x$, where
MC statistics is insufficiently good, we have taken into account
that at small $x$ the function $u(x)$
can be expanded in powers of $x$.
In the limit of strong Coulomb coupling ($\Gamma \gg 1$)
only even powers of $x$ should survive \cite{widom63}
and the expansion should
have the form
\begin{equation}
     u(x)=\alpha_0 + \alpha_2 \,x^2 +
     \alpha_4\,x^4+\alpha_6\,x^6+ \ldots,
\label{widom}
\end{equation}
where $\alpha_0=h_0(\Gamma)/\Gamma$, $\alpha_2=-{1 \over 4}$
\cite{jancovici77}, and
$\alpha_4$, $\alpha_6$, \ldots can depend on $\Gamma$. Thus,
$\alpha_0$ is a slowly varying function of $\Gamma$
determined by the function $h_0(\Gamma)$ (discussed
in Sec.\ \ref{sect-TS}). One can use any accurate
representation of $h_0(\Gamma)$, for instance, Eqs.\ (a) or (b)
from Table \ref{tab:h0}. However, the
fit formula (\ref{hfit}) for $u(x)$ presented below
is especially accurate if $h_0(\Gamma)$ is
given by our own fit expression (\ref{krokodil}) (see Sec.\ \ref{sect-fit}).
For a not too strong Coulomb coupling
($\Gamma\lesssim 5$),
odd powers of $x$ can become pronounced in the expansion (\ref{widom}),
and %$\alpha_0$ and
$\alpha_2$ can be modified.

We have fitted our MC data points by the analytic expression
\begin{equation}
    u(x)=\alpha_0\left[
                    \frac{
                          1-C_4\,x-2\,(C_1/\alpha_0)\,x^2+C_3\,x^4+C_2\,x^8
                         }
                         {
                          1+C_2\,\alpha_0^2\,x^{10}
                         }
                \right]^{1/2},
\label{hfit}
\end{equation}
where
\begin{eqnarray}
    C_1&=&0.25-0.267\,\Gamma^{-1.44},
%\nonumber   \\
\quad
    C_2=0.05,
\nonumber   \\
    C_3&=&0.084-0.144\,\Gamma^{-1.7},
%\nonumber  \\
\quad
    C_4= 0.434\,\Gamma^{-1.2}.
\label{Afit}
\end{eqnarray}

The fit quality is demonstrated in Fig.\ \ref{fig:potential}. The root-mean-square
relative error is 0.4\%, the maximum error 1.1\% takes place at $\Gamma=10$ and
$x=1.67$. It is seen that the fit is accurate for a classical Coulomb liquid (any
$\Gamma \ge 1$) at least at $x \lesssim 3$. This is sufficient to calculate the plasma
screening enhancement of reaction rates (Sec.\ \ref{sect-enhancement}). It is remarkable
that the fit is valid also in the crystalline solid ($\Gamma=400$ and 800) although we
have not included the crystalline data into the fitting. Clearly, $H(r)$ in the solid
and strongly coupled liquid ($\Gamma \gg 1$) is nearly the same, being ``frozen''
(almost independent of $\Gamma$). This allows us to expect that Coulomb tunneling and
the fusion reaction rate should not undergo significant changes when the temperature
drops below the freezing temperature $T_m$.

%%%%%%%%%%%%%%%%%%%%%%%%%%%%%%%%%%%%%%%%%%%%%%%%%%%%%
\subsection{Enhancement factor}
\label{sect-enhancement}
%%%%%%%%%%%%%%%%%%%%%%%%%%%%%%%%%%%%%%%%%%%%%%%%%%%%%

Having $H(r)$ we can introduce the
mean-field reaction rate
\begin{eqnarray}
  R^\mathrm{MF}&=&\frac{n_i^2 S_\mathrm{pk}}{2}\,  %\,
  \sqrt{8  \over \pi\mu (k_BT)^3}\,
  %\int_{E\mathrm{min}}^\infty {\rm d} E\,
\nonumber \\
  && \times \int_{E\mathrm{min}}^\infty {\rm d} E\,
  \exp\left[-\frac{E}{k_B T} -P(E) \right],
\label{thermon}
\end{eqnarray}
where $E$ is an energy of relative motion of colliding nuclei
(with a minimum value $E_\mathrm{min}$ at the bottom of the potential well),
$\exp(-E/k_BT)$ comes from the Maxwellian energy distribution
of the nuclei, and
$S_\mathrm{pk}$ is the $S$-factor corresponding to the energy $E$
at which the integrand has maximum. Finally,
\begin{equation}
     P(E)={ 2 \sqrt{2\mu} \over \hbar} \,
     \int_{r_n}^{r_t} {\rm d}r \,
     \sqrt{{Z^2e^2 \over r} -H(r) -E}
\label{P(E)}
\end{equation}
characterizes the penetrability of the Coulomb barrier at an energy $E$; $r_n$ and $r_t$
are classical turning points which are zeros of the expression under the square root; we
can set $r_n \to 0$ (an exact $r_n$ should have been determined by nuclear interactions
which we neglect in the Coulomb tunneling problem). Here we use a radial WKB
approximation with the mean-field potential $H(r)$. A similar approach was used by Ogata
{\it et al.}\ \cite{oii91,oiv93}. The main difference is that Ogata {\it et al.}\
numerically solved a radial Schr\"odinger equation which should be equivalent to the WKB
integration under the conditions of study. Notice that Ogata {\it et al.}\ employed less
accurate mean-field potentials and obtained, therefore, less accurate results as
discussed e.g., in \cite{rosenfeld96,leandro05} (also see Sec.\ \ref{sect-TS}). The
approach equivalent to our but with a less accurate mean-field potential was used by
Itoh {\it et al}.\ \cite{ikm90} (Sec.\ \ref{sect-TS}).

Putting $H(r)=0$ in Eq.\ (\ref{P(E)}) we reproduce the well known result
$P(E)=2 \pi Z^2e^2/(\hbar v)$ for a pure Coulomb
barrier (with $v=\sqrt{2E/\mu}$).
Taking energy integral in Eq.\ (\ref{thermon})
by a saddle-point method
we come to the thermonuclear reaction rate
without plasma screening, $R^\mathrm{MF}\{H=0\}=R_\mathrm{th}$,
given by Eq.\ (\ref{therm}).

In the WKB mean-field approximation the
enhancement factor of the nuclear reaction rate is
\begin{equation}
   F_\mathrm{scr}^\mathrm{MF}=R^\mathrm{MF}\{H\}/R^\mathrm{MF}\{0\}.
\label{WKBenh}
\end{equation}
The calculation of
$F_\mathrm{scr}^\mathrm{MF}$ from Eq.\ (\ref{WKBenh})
reduces to the evaluation of $P(E)$ and
$R^\mathrm{MF}\{H\}$ from Eqs.\ (\ref{P(E)})
and (\ref{thermon}) for a given $H(r)$. We have performed the integrations
over $r$ and $E$ numerically (beyond the saddle-point approximation)
in all integrals.

These results are naturally restricted by the WKB
and mean-field approximations.
They neglect fluctuative nature of plasma microfields;
deviations from spherical symmetry and dynamical evolution of
these microfields during a tunneling event; corrections to $H(r)$
due to quantum effects in ion motion; deviations from the
first-order one-dimensional (radial) WKB approximation.
In the thermonuclear regime with strong plasma screening
all these effects are not expected to be strong.

%%%%%%%%%%%%%%%%%%%%%%%%%%%%%%%%%%%%%%%%%%%%%%%%%%%%%%%%%%%%%%%%%%%%%%%%
\subsection{Analytic fit}
\label{sect-fit}
%%%%%%%%%%%%%%%%%%%%%%%%%%%%%%%%%%%%%%%%%%%%%%%%%%%%%%%%%%%%%%%%%%%%%%%%

To facilitate applications of our results we have fitted
the values of the enhancement factor
$F_\mathrm{scr}^\mathrm{MF}$, calculated from Eq.\ (\ref{WKBenh}),
by an analytic expression. First of all, we notice that
at $1 \leq \Gamma \leq 200$ the function $h_0(\Gamma)$
can be approximated as
\begin{eqnarray}
   h_0^\mathrm{fit}(\Gamma)&=&\Gamma^{3/2}
   \left( { A_1 \over \sqrt{A_2+\Gamma}} +{ A_3 \over 1+ \Gamma} \right)
   \nonumber \\
   &&+{B_1 \Gamma^2 \over B_2 + \Gamma}+{B_3 \Gamma^2 \over B_4+\Gamma^2},
\label{krokodil}
\end{eqnarray}
with $A_1=2.7822$, $A_2=98.34$, $A_3=\sqrt{3}-A_1/\sqrt{A_2}=1.4515$,
$B_1=-1.7476$, $B_2=66.07$, $B_3=1.12$, and $B_4=65$.
The accuracy of this fit is the same as the accuracy of
the best fit (b) in Table \ref{tab:h0}.
At $\Gamma \gg 1$ it gives
$h_0^\mathrm{fit}(\Gamma) \approx 1.0346 \, \Gamma$
and remains accurate at $200 \leq \Gamma \leq 600$.
In addition, it reproduces the correct
Debye-H\"uckel asymptote
$h_0^\mathrm{fit}(\Gamma) = \sqrt{3}\Gamma^{3/2}$
at $\Gamma \ll 1$.
%DGY inserted
Note that the functional form of Eq.\ (\ref{krokodil}) was suggested in Ref.\
\cite{pc00} to approximate the free energy of OCP.

We have fitted
the values of the enhancement factor,
calculated on a dense grid of values
$1 \leq \Gamma \leq 200$ and $0 \leq \zeta \leq 8$,
as $F_\mathrm{scr}^\mathrm{MF}
=\exp(h_\mathrm{MF}^\mathrm{fit})$,
\begin{equation}
  h_\mathrm{MF}^\mathrm{fit}(\Gamma,\zeta)=h_0^\mathrm{fit}(\Gamma)+
  h_1^\mathrm{fit}(\Gamma,\zeta)=h_0^\mathrm{fit}(\widetilde{\Gamma}),
\label{hifit}
\end{equation}
with
\begin{equation}
  \widetilde{\Gamma}=\Gamma/
  (1+\alpha \zeta + \beta \zeta^2 + \gamma \zeta^3)^{1/3},
\label{Gammafit}
\end{equation}
$\alpha=0.022$, $\beta=0.41-0.6/\Gamma$, and $\gamma=0.06+2.2/\Gamma$.
The maximum fit error of $F_\mathrm{scr}^\mathrm{MF}$
is $\approx 30\%$; it occurs at $\zeta=0.4$ and $\Gamma=200$,
at which the enhancement factor itself
is enormously large,  $F_\mathrm{scr}^\mathrm{MF}\sim 10^{90}$.
The fit accuracy is illustrated in Fig.\ \ref{fig:etagamma}
where the dashed lines (right vertical scales)
show the ratio of fitted and calculated WKB mean-field
values of $g_\mathrm{MF}(0)$ [same as the ratios of fitted and calculated
values of $F_\mathrm{scr}^\mathrm{MF}$].
We have checked that an extension of $\zeta$ from 8 to 50
at $1 \leq \Gamma \leq 200$ does not change
the initial fit accuracy. Moreover, if $\zeta \leq 50$,
the fit remains sufficiently accurate up to $\Gamma \sim 600$.
For instance, at $\Gamma=400$ the fit gives
the values of $F_\mathrm{scr}^\mathrm{MF}$ which differ from
the calculated values within a factor of 2, and at $\Gamma=600$ --
within a factor of 10. If $\Gamma \gg 1$ and
$\zeta \lesssim 1$, the main difference between $\widetilde{\Gamma}$
and $\Gamma$ in Eq.\ (\ref{Gammafit})
is determined by the term in the denominator
containing $\beta \approx 0.41$ (while the terms
containing $\alpha$ and $\gamma$ are relatively small).
Neglecting $\alpha$ and $\gamma$ for a moment and
treating $\beta \zeta^2$ as a small correction, we
obtain $h_1\approx -1.0346 \times 0.41\Gamma \zeta^2/3
\approx -0.141 \Gamma \zeta^2$, which is very close
to $-(5/32)\Gamma \zeta^2 \approx -0.156 \Gamma \zeta^2$
given by Eq.\ (A) of Table \ref{tab:h1} and discussed in
Sec.\ \ref{sect-TS}.

To summarize, our fit gives
very accurate values of $F_\mathrm{scr}^\mathrm{MF}$
for all possible values of $\Gamma$ and $\zeta$ at
which the WKB mean-field approximation is valid
(see below) and, actually, in much wider domain.

%%%%%%%%%%%%%%%%%%%%%%%%%%%%%%%%%%%%%%%%%%%%%%%%%%%%%%%%%%%%%%%%%%%%%%%%%
\section{PIMC, mean-field WKB, and other results}
\label{sect-analysis}
%%%%%%%%%%%%%%%%%%%%%%%%%%%%%%%%%%%%%%%%%%%%%%%%%%%%%%%%%%%%%%%%%%%%%%%%%

%
\begin{figure}[t!]
    \begin{center}
        \leavevmode
        \epsfxsize=3.2in
    \epsfbox[10 15 387 790]{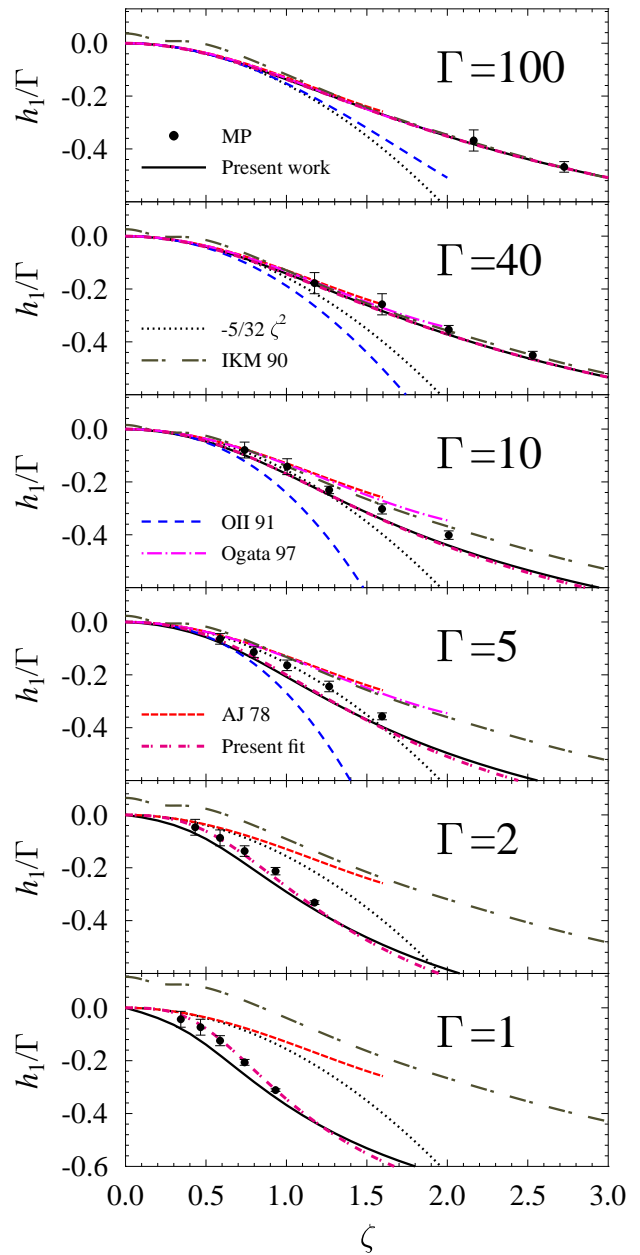}%{h1.eps}
    \end{center}
    \caption{(Color online)
    The screening function $h_1^\mathrm{MF}(\Gamma,\zeta)/\Gamma$
    versus $\zeta$ at six values of $\Gamma$.
    Dots with errorbars
    are Militzer-Pollock points [of group (ii)
    in Table \ref{tab:mp}], solid lines are our mean-field
    WKB calculations, and short dash-dot-space lines
    are given by our fit. Dotted lines show the lowest-order
    result of Jancovici \cite{jancovici77}, while
    short-dash,
    long dash-dot-space,
    long-dash, and  dot-dash lines present the results of Alastuey and
    Jancovici (AJ 78, Ref.\ \cite{aj78}),
    Itoh {\it et al.}\ (IKM 90, Ref.\ \cite{ikm90}),
    Ogata {\it et al.}\
    (OII 91, Ref.\ \cite{oii91}),
    and Ogata (Ogata 97, Ref.\ \cite{ogata97}), respectively.
    }
    \label{fig:h1}
\end{figure}
%

%%%%%%%%%%%%%%%%%%%%%%%%%%%%%%%%%%%%%%%
\subsection{Overall analysis}
\label{sect-overall}
%%%%%%%%%%%%%%%%%%%%%%%%%%%%%%%%%%%%%%%

After calculating $F_\mathrm{scr}^\mathrm{MF}$
we have used Eq.\ (\ref{klevo}) and determined $g_\mathrm{MF}(0)$.
The results are shown by solid lines in Fig.\ \ref{fig:etagamma}
for the same five values of $\eta$ which were taken by
Militzer and Pollock \cite{mp05}. This allows us to
directly compare the PIMC and mean-field WKB approaches
for all PIMC points [all groups (i)--(iii) of data points in Table
\ref{tab:mp}].
Open symbols (right vertical scale) show ratios of
the Militzer-Pollock to calculated
mean-field results. The errorbars are those as reported  \cite{mp05}
in the PIMC simulations  of $g_{MP}(0)$
(with the corrections mentioned in Sec.\ \ref{sect-mp}).
The overall agreement seems very satisfactory.
Large differences take place in pycnonuclear points.
Very strong differences $g_{MP}(0)/g_\mathrm{MF}(0)\sim
0.04$ and $\sim 10^{-4}$ occur
for $\eta=1$ at $\Gamma=400$ and 600, respectively.
%(as indicated by arrows in Fig.\ \ref{fig:etagamma}).
We analyze all these results below.

%%%%%%%%%%%%%%%%%%%%%%%%%%%%%%%%%%%
\subsection{Data of group (ii)}
\label{groupii}
%%%%%%%%%%%%%%%%%%%%%%%%%%%%%%%%%%%

After calculating the enhancement
factor $F_\mathrm{scr}^\mathrm{MF}$,
we have presented it in the form
(\ref{Fscr}) and determined $h^\mathrm{MF}$.
Using then Eqs.\ (\ref{h0h1}) and
(\ref{krokodil})
%Eq.\ (b) of Table \ref{tab:h0}
we have calculated
$h_1^\mathrm{MF}=h^\mathrm{MF}-h_0(\Gamma)$. This function
is not very certain and has been a subject
of debates (Secs.\ \ref{sect-TS} and \ref{sect-all}).
Solid lines in Fig.\ \ref{fig:h1} show
our calculated values of $h_1(\Gamma,\zeta)/\Gamma$
versus $\zeta$ for six values of $\Gamma=1$, 2, 5, 10, 40, and 100.
Short dash-dot-space lines are given by our fit expression
(Sec.\ \ref{sect-fit}).

Furthermore, taking the contact probabilities
calculated by Militzer and Pollock in the
points which belong to group (ii) (Table \ref{tab:mp})
and using Eq.\ (\ref{klevo}) we have determined
the PIMC values of $h_1(\Gamma,\zeta)$
for the same six values of $\Gamma$ as in Fig.\
\ref{fig:h1} at several values of $\zeta$.
These data are plotted in Fig.\ \ref{fig:h1} by dots,
together with numerical errorbars of Militzer and
Pollock \cite{mp05}. % (also see Sec.\ \ref{sect-mp}).
One can observe a remarkably good
agreement between the mean-field WKB and PIMC results for the
data of group (ii).
Strongest disagreement
occurs at the lowest $\Gamma=1$, where the function
$h_1$ introduces small contribution into
the plasma screening enhancement of the nuclear reaction rates.
The existence of real disagreement at $\Gamma \sim 1$
between the PIMC and the mean-field WKB results
could be checked in future
more extensive PIMC runs. If real, this disagreement could be attributed
to a fluctuative nature of the plasma potential
at $\Gamma\lesssim 1$ (where Coulomb coupling
is not too strong and can allow noticeable fluctuations
of the plasma potential from its mean-field values).
This would indicate that
the PIMC results are more accurate at $\Gamma \lesssim 1$
than the WKB mean-field results.

\begin{figure}
    \begin{center}
        \leavevmode
        \epsfxsize=3.05in
    \epsfbox[40 30 275 655]{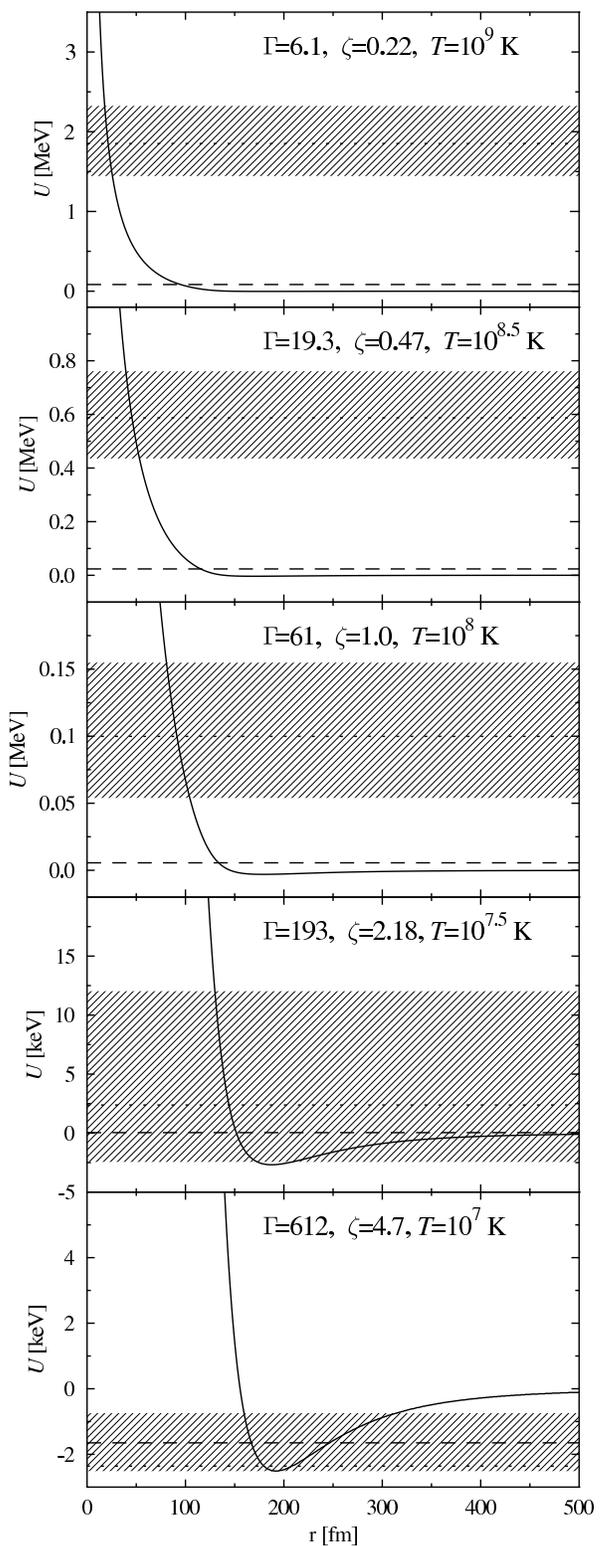}  %{gamow.eps}
    \end{center}
    \caption{Effective mean-field Coulomb
    potential $U(r)$ for reacting carbon nuclei
    at $\rho=5 \times 10^9$ g~cm$^{-3}$ and
    five temperatures ($\log_{10} T[\mathrm{K}]
    =9,$ 8.5, 8, 7.5, and 7). Shaded strips show
    Gamow-peak energy ranges; dotted lines are
    Gamow-peak energies; dashed lines are thermal
    energies $k_BT$ measured from the bottom of $U(r)$.
    }
    \label{fig:gamow}
\end{figure}

We have also compared the Militzer-Pollock and our results with some other calculations
of $h_1(\Gamma,\zeta)$ discussed in Sec.\ \ref{sect-TS}. In particular, the dotted lines
in Fig.\ \ref{fig:h1} show the basic lowest-order expression (A) from Table \ref{tab:h1}
obtained by Jancovici \cite{jancovici77}; the short-dashed, long-dashed, and dot-dashed
lines display the expressions (B), (C), and (D) calculated, respectively, by Alastuey
and Jancovici \cite{aj78}, Ogata {\it et al.} \cite{oii91}, and Ogata \cite{ogata97}.
These results are shown for those ranges of $\Gamma$ and $\zeta$ for which they were
obtained in the cited publications (Table \ref{tab:h1}). In addition,
long-dash-dot-space lines present the function $h_1(\Gamma,\zeta)$ which corresponds to
the results of Itoh {\it et al}.\ \cite{ikm90} (discussed in Sec.\ \ref{sect-TS}). We
see that the Militzer-Pollock and our results are in good agreement with earlier
predictions of Jancovici, and especially of Alastuey and Jancovici, and Ogata
\cite{ogata97}, but they are in worse agreement with the results of Ref.\ \cite{oii91}.
% at not
%too small $\zeta$.
Calculations of Ref.\ \cite{ikm90}
are also accurate; some whirls of corresponding curves
at small $\zeta$ occur possibly because of simplified approximation
of the mean potential in the cited publication
(see Sec.\ \ref{sect-TS}).

\begin{figure}
    \begin{center}
        \leavevmode
        \epsfxsize=75mm
    \epsfbox{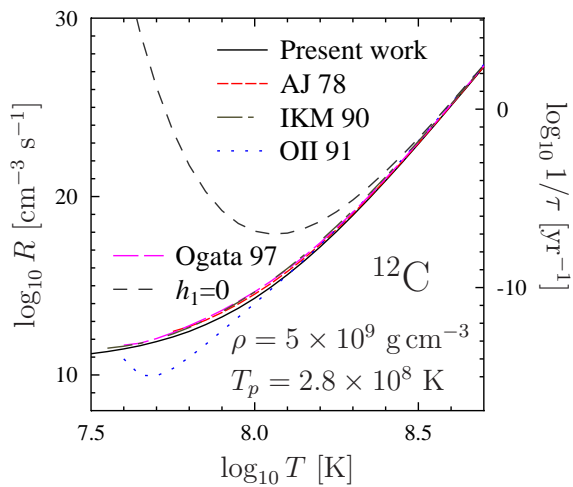} %{temper.eps}
    \end{center}
    \caption{(Color online) Temperature dependence of
    the carbon burning rate $R$
    (left vertical scale) and the inverse carbon burning time
    $\tau^{-1}=R/n_i$ (right vertical
    scale) at $\rho=5\times 10^9$ g~cm$^{-3}$
    for the temperature range $T \gtrsim T_p$ of thermonuclear
    burning with strong screening and for lower $T$ (down
    to $T \sim 0.1 T_p$).
    The upper line is obtained with the screening enhancement function
    $h=h_0(\Gamma)$ given by Eq.\ (b) in Table \ref{tab:h0}.
    Other lines employ the total function $h=h_0+h_1$ calculated
    by Alastuey and Jancovici  (AJ 78, Ref.\ \cite{aj78}),
    Itoh {\it et al}.\ (IKM 90, Ref.\ \cite{ikm90}),
    Ogata {\it et al.}\  (OII 91, Ref.\ \cite{oii91}), and Ogata
     (Ogata 97, Ref.\ \cite{ogata97}). The solid line plots
    our WKB mean-field calculation. See text for details.
    }
    \label{fig:temper}
\end{figure}

As a byproduct of our calculations in the mean-field WKB approximation,
we have determined the energy $E_\mathrm{pk}$
of the peak of the integrand function
in Eq.\ (\ref{thermon}); it is the
Gamow-peak energy modified by the plasma screening effects.
We have also estimated characteristic energy widths of the Gamow
peak (at the half width of the peak maximum).
In Fig.\ \ref{fig:gamow} we show the effective total radial mean-field
Coulomb potential $U(r)=Z^2e^2/r-H(r)$ for  $^{12}$C ions
reacting in pure carbon matter at $\rho=5 \times 10^9$ g~cm$^{-3}$.
At this $\rho$, the ion-sphere radius is $a$=98 fm. Naturally,
$U(r)$ has a minimum at $r \approx 2a$ due to Coulomb coupling.
We plot $U(r)$ for five temperatures,
$T=10^9$, $10^{8.5}$, $10^8$, $10^{7.5}$, and $10^7$~K.
With decreasing $T$, the minimum becomes more
pronounced and finally ``freezes'' at $\Gamma \gtrsim 100$.
The shaded strips in Fig.\ \ref{fig:gamow} show the Gamow-peak
energy ranges and the dotted lines show $E_\mathrm{pk}$.
The dashed lines present the thermal energy level $k_BT$
measured from the bottom of $U(r)$.

The first two upper panels in Fig.\ \ref{fig:gamow}
refer to the thermonuclear reaction regime with strong plasma
screening ($T \gtrsim T_p$). The next two panels are for
a colder plasma ($T=0.368 T_p$ and $0.114 T_p$,
respectively), while the lowest panel is for a very cold
plasma ($T=0.0361 T_p$), certainly in the zero-temperature
pycnonuclear regime. When the temperature decreases,
the Gamow-peak energy range becomes thinner
(note the difference of energy scales
in different panels!) and shrinks to lower energies,
together with $E_\mathrm{pk}$.
In the three upper panels the Gamow peak range is
still at $E>0$ [belonging to continuum states in a potential $U(r)$].
The energies from this range are much higher than $k_BT$
supporting the statement that the main contribution into
reaction rates at sufficiently high $T$ comes from suprathermal ions.
In these cases, the underlying mean-field WKB approximation
can be adequate.
In the forth panel, the lowest energies of the Gamow-peak range
become negative (drop to bound states) although $E_\mathrm{pk}$
is still positive and higher than $k_BT$. The mean-field
WKB approach may be qualitatively correct but
quantitatively inaccurate. At the bottom panel
the Gamow-peak energy range fully shrinks to
bound-state energies and the formal Gamow-peak energy becomes
lower than $k_BT$. It is clear that the mean-field WKB
approximation breaks down at these low temperatures,
and the formally calculated $E_\mathrm{pk}$ is inaccurate.
Therefore, it is natural that the mean-field WKB results
diverge from the PIMC ones at low temperatures.
This divergence is seen in Fig.\ \ref{fig:etagamma}
(at highest values of $\Gamma$, especially at $\Gamma=400$ and
600 for $\eta=1$).

%Let us remark that the values of $E_\mathrm{pk}$ discussed above
%refer to relative motion of the reacting nuclei before
%Coulomb tunneling. After the tunneling in a strongly
%coupled plasma ($\Gamma \gtrsim 1$), this energy increases
%by $\approx k_BT h_0(\Gamma)$ (see, e.g., Ref.\ \cite{itw03})
%due to Coulomb interaction
%in a plasma environment (and
%appears to be always positive even if $E_\mathrm{pk}$ becomes negative).
%This increased energy determines the efficiency
%of the nuclear reaction after the Coulomb penetration and
%should be substituted into
%the argument of the $S$-factor.

As follows from the above consideration, the mean-field WKB approximation
can be valid for $\zeta \lesssim 1.6-3$
[$T \gtrsim (0.1-0.2) T_p$]. This is further illustrated
in Fig.\ \ref{fig:temper} which shows the carbon burning
rate versus temperature at the same density
$\rho=5 \times 10^9$ g~cm$^{-3}$ as in Fig.\ \ref{fig:gamow}.
The upper line is obtained with the simplified
enhancement factor $F_\mathrm{scr}=\exp[h_0(\Gamma)]$,
where $h_0(\Gamma)$ is given by Eq.\ (b) in Table \ref{tab:h0}.
It is seen to be a good approximation in the thermonuclear
regime ($T \gtrsim T_p$) but gives qualitatively
wrong results just after $T$ decreases below $T_p$.
It is easy to check that in the thermonuclear regime
with strong screening
$h_1$ is indeed a small correction to $h_0$.

However, adding $h_1$ and using
$F_\mathrm{scr}=\exp(h_0+h_1)$ greatly helps
extending the strong-screening thermonuclear results to lower temperatures,
down to $T \sim (0.1-0.2)T_p$. Below $T_p$ the function
$h_1$ is no longer small but becomes comparable to $h_0$ and
crucial to get physically reasonable results.
All other lines in
Fig.\ \ref{fig:temper} are plotted by adding $h_1$
calculated in the various approximations. The solid line
shows our mean-field WKB calculations which are nearly identical
to our fit and
to the Militzer-Pollock PIMC results in the displayed
temperature range. The short-dashed line corresponds to
$h_1$ and $h_0$ obtained by Jancovici \cite{jancovici77}
and Alastuey and Jancovici \cite{aj78}  for $\zeta \leq 1.6$
[Eqs.\ (b) and (B)
in Tables \ref{tab:h0} and \ref{tab:h1}].
%It is in a very good agreement with the Militzer-Pollock and
%our results.
The dash-dot-space line is plotted using the fit formula
of Itoh {\it et al}.\ \cite{ikm90} ($\zeta \leq 5.4$).
%which also agrees well
%with these results.
The dotted line in Fig.\ \ref{fig:temper} shows the results of
Ogata {\it et al.} \cite{oii91,oiv93}
($\zeta \lesssim 2$, Tables \ref{tab:h0} and \ref{tab:h1})
and the long-dashed line shows the PIMC calculations
of Ogata \cite{ogata97}
(also $\zeta \lesssim 2$, the same tables).
We see that adding $h_1$ makes the reaction rate
at $T \sim 0.1 T_p$ almost temperature independent,
as it should be in the pycnonuclear regime.

It is remarkable that all cited results
(except for Refs.\ \cite{oii91,oiv93}) obtained
at $T \gtrsim (0.1-0.2)\,T_p$
in different techniques and using various simplified
assumptions (Sec.\ \ref{sect-TS}) give actually
{\it almost one and the same} reaction rate
(almost the same curve in Fig.\ \ref{fig:temper}) reproduced by the
mean-field WKB approach. It gives us confidence that
this approach is really valid at $T \gtrsim (0.1-0.2)\,T_p$,
and can now be considered as very reliable.
It has been expected by many authors (e.g., Ref.\ \cite{ikm90}),
and it is strictly confirmed by the Militzer-Pollock
PIMC calculations \cite{mp05}.

%At $T \lesssim 0.3 T_p$ both curves disagree with
%each other and with Alastuey-Jancovici,
%Itoh {\it et al.}, Militzer-Pollock
%and our curves. The nature of disagreement is discussed
%in Sec.\ \ref{sect-TS}. We expect that the Alastuey-Jancovici,
%Itoh {\it et al}., Militzer-Pollock and our results are more accurate than
%the results of other authors and reliably extend
%the expressions, obtained in the thermonuclear regime,
%to temperatures somewhat below $T_p$.
% down
%to $T \sim (0.1-0.2)T_p$, i.e., through the intermediate
%thermo-pycnonuclear burning regime and through the highest
%temperatures where the burning is already pycnonuclear.

%%%%%%%%%%%%%%%%%%%%%%%%%%%%%%%%%%%%%%%%%%%%%%%%
\subsection{Pycnonuclear Militzer-Pollock data [group (iii)]}
\label{sect-pycno-mp}
%%%%%%%%%%%%%%%%%%%%%%%%%%%%%%%%%%%%%%%%%%%%%%%%

%
\begin{figure}
    \begin{center}
        \leavevmode
        \epsfxsize=3.2in \epsfbox{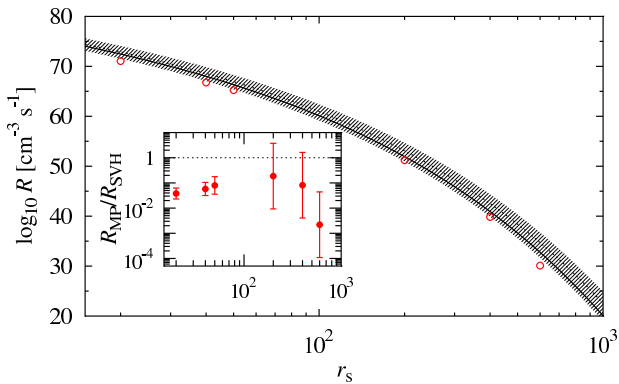}
    %{PycnReact2_corrected2.eps}
    \end{center}
    \caption{Pycnonuclear temperature-independent
    carbon burning rate
    versus the density parameter $r_s$.
    Open dots show Militzer-Pollock points. Shaded strip
    presents theoretical uncertainties of other calculations
    and solid line is the optimal model
    among these calculations \cite{leandro05}. Filled circles
    with errorbars in the inset give the ratio of
    the Militzer-Pollock rates to the predictions of
    the optimal (Salpeter and Van Horn, SVH) model.
    }
    \label{fig:pycnos}
\end{figure}

Finally let us analyze six Militzer-Pollock data
points which correspond to temperature-independent
pycnonuclear burning [the data of group (iii) in Table \ref{tab:mp}].
In Fig.\ \ref{fig:pycnos}
we plot
the zero-temperature pycnonuclear carbon burning rate
as a function of the density parameter $r_s$ defined by Eq.\ (\ref{rs}).
For simplicity, in this
figure we use a constant (energy independent)
astrophysical factor $S(E=1~\mathrm{MeV})
=3.2\times10^{16}$ barn~MeV$^{-1}$
for calculating the reaction rates.
The six Militzer-Pollock reaction rates are shown by open dots.

We have compared these data with theoretical calculations
of other authors. % discussed in Sec.\ \ref{sect-zeropyc}.
The shaded strip shows other theoretical predictions
taking into account the uncertainties of various approximations.
It is plotted using the expressions of
Gasques {\it et al.}\ \cite{leandro05}
who analyzed calculations of different authors. The strip is restricted
by the minimum and maximum allowable reaction rates
suggested in \cite{leandro05}.
The solid line is the optimal theoretical reaction rate, which
is the static-lattice model of Salpeter and Van Horn \cite{svh69}
for a bcc Coulomb crystal.

The consistency of the Militzer-Pollock data with other results is satisfactory. One
should take in mind that the PIMC calculations in the pycnonuclear regime can be not too
accurate because they require best computer resources. Moreover, the theoretical
predictions of Ref.\ \cite{leandro05} refer to pycnonuclear reactions in Coulomb crystal
while the actual state of ions in the PIMC runs is unknown (not reported in
\cite{mp05}). As pointed out by many authors (see, e.g., Refs.\
\cite{ca80,chabrier93,jc96}), strong zero-point vibrations of ions at $r_s \gtrsim
90-160$ prevent their crystallization even at $T=0$. Therefore, three low-$r_s$ points
of Militzer and Pollock in Fig.\ \ref{fig:pycnos} should correspond to pycnonuclear
burning in cold quantum liquid. However they do not deviate strongly from the
predictions for Coulomb crystals.

Nevertheless, the agreement of the pycnonuclear Militzer-Pollock
points with the best theoretical prediction of Salpeter and
Van Horn \cite{svh69} is not perfect. Filled dots
in the inset in Fig.\ \ref{fig:pycnos} display the ratio
of the Militzer-Pollock to the best Salpeter-Van Horn reaction rates;
the errorbars are those reported in \cite{mp05}
(with the corrections mentioned in Sec.\ \ref{sect-mp}). We see
that the Militzer-Pollock rates are noticeably
lower. The nature of this difference is unknown.

We expect that the Militzer-Pollock calculations in the
pycnonuclear regime
are not superior over other theoretical predictions at
low $T$.
More extended PIMC
studies in the pycnonuclear regime would be helpful
to reduce current theoretical uncertainties of the reaction
rates.

%%%%%%%%%%%%%%%%%%%%%%%%%%%%%%%%%%%%%%%%%%%%%%%%%%
\section{Conclusions}
\label{sect-conclusions}
%%%%%%%%%%%%%%%%%%%%%%%%%%%%%%%%%%%%%%%%%%%%%%%%%%

We have analyzed the recent Path Integral
Monte Carlo (PIMC) calculations
by Militzer and Pollock \cite{mp05}
of contact probabilities of atomic nuclei
participating in fusion reactions
in dense matter.
We have compared these calculations with other
theoretical predictions.
In particular, we have used
a simple model based on WKB radial Coulomb tunneling of
the reacting nuclei in the static mean-field plasma potential
created by plasma ions.
We have employed accurate Monte Carlo (MC) calculations
of the mean-field plasma potential for a one-component strongly
coupled plasma of ions and proposed a simple and accurate analytic fit to
the plasma potential (Sec.\ \ref{sect-mfpot}).

Our main conclusions are as follows:
\begin{enumerate}

\item
We have found a very good agreement of the
Militzer-Pollock PIMC results with the mean-field WKB calculations
for $T \gtrsim (0.1-0.2) T_p$, i.e., in the thermonuclear regime,
intermediate thermo-pycnonuclear regime and at highest
temperatures of pycnonuclear burning (Table \ref{tab:regimes}).
These results show good agreement with theoretical predictions of
many authors (e.g., \cite{jancovici77,aj78,ikm90,ogata97})
and can be considered as well established.

\item
There is a tentative slight disagreement of the PIMC
and mean-field WKB results in the case of moderately
strong ion coupling $\Gamma \lesssim 1$ but it cannot
strongly affect the reaction rates.

\item
We have obtained a very accurate fit (Sec.\ \ref{sect-fit})
to the plasma screening enhancement factors calculated in the
mean-field WKB approximation. The fit reliably describes
the PIMC results
in a wide temperature range $T \gtrsim (0.1-0.2)T_p$.

\item
New studies of Coulomb tunneling problem in the pycnonuclear
regime are needed to obtain accurate reaction rates in this regime.

\end{enumerate}

The validity of the mean-field
WKB method at $\Gamma \gtrsim 1$ and $T\gtrsim T_p$ could be
expected (strong Coulomb coupling arranges quasi-order
which may be well described by a radial mean-field
without fluctuations).
In contrast, there is little doubt that at $T \ll T_p$
the radial mean-field
WKB picture is not true. At these low
temperatures the reacting ions occupy
quantum energy levels in their potential wells;
they fuse along selected (anisotropic) close-approach trajectories
\cite{svh69} which is definitely beyond the mean-field
radial WKB method.
Because the accuracy of PIMC calculations
\cite{mp05} decreases at low $T$,
a nice agreement between the PIMC \cite{mp05} and mean-field
WKB approaches at $T \sim 0.1 T_p$, which we formally reached,
may indicate insufficient accuracy of the low-$T$ PIMC results.
New PIMC simulations would be most desirable to clarify this point.

\begin{acknowledgments}
%DGY included AYP
We are grateful to B.~Militzer, V.~K.\ Nikulin, and A.~Y.\ Potekhin for useful remarks.
%We are grateful to B.~Militzer and V.~K. Nikulin for useful remarks.
Work of AIC and DGY was partly supported by the
Russian Foundation for Basic Research (grants 05-02-16245,
05-02-22003), by the Federal Agency for Science and Innovations
(grant NSh 9879.2006.2), and by the Dynasty Foundation. Work of HED
was performed under the auspices of the US Department of Energy by
the Lawrence Livermore  National Laboratory under contract number
W-7405-ENG-48.
\end{acknowledgments}

%%%%%%%%%%%%%%%%%%%%%%%%%%%%%%%%%%%%%%%%%%%%%%%%%%%%%%%%%%%%%%%%%%%%%%%%%%%%%%%
                    
%%%%%%%%%%%%%%%%%%%%%%%%%%%%%%%%%%%%%%%%%%%%%%%%%%%%%%%%%%%%%%%%%%%%%%%%%%%%%%%
\end{document}